\documentclass[man,floatsintext]{apa7}
\usepackage{style}

\authorsnames[1,2,3]{Hau-Hung Yang,
Chia-Min Wei, Yu-Chang Chen}
\authorsaffiliations{{Department of Psychology, National Taiwan University},{Department of Economics, University of Wisconsin-Madison},{Department of Economics, National Taiwan University}}

\shorttitle{Bayesian Adaptive Testing}


\abstract{\justifying This study establishes the consistency of Bayesian adaptive testing methods under the Rasch model, addressing a gap in the literature on their large-sample guarantees. Although Bayesian approaches are recognized for their finite-sample performance and capability to circumvent issues such as the cold-start problem; however, rigorous proofs of their asymptotic properties, particularly in non-i.i.d. structures, remain lacking. 
	We derive conditions under which the posterior distributions of latent traits converge to the true values for a sequence of given items, and demonstrate that Bayesian estimators remain robust under the mis-specification of the prior. Our analysis then extends to adaptive item selection methods in which items are chosen endogenously during the test. Additionally, we develop a Bayesian decision-theoretical framework for the item selection problem and propose a novel selection that aligns the test process with optimal estimator performance. These theoretical results provide a foundation for Bayesian methods in adaptive testing, complementing prior evidence of their finite-sample advantages.}

\date{\today}

\usepackage[style=apa,sortcites=true,sorting=nyt,backend=biber]{biblatex}
\authornote{
   \addORCIDlink{Hau-Hung Yang}{0009-0009-3318-4110} 
    \addORCIDlink{Yu-Chang Chen}{0000-0003-3851-6902} 

}
\begin{document}
\title{On the Consistency of Bayesian Adaptive Testing under the Rasch Model}
\maketitle	

\justifying

\setcounter{secnumdepth}{3}

\section{Introduction}

Item Response Theory (IRT) is widely used in psychometrics to construct measurement models and effectively measure abilities \parencite{Ayala2022}. Many studies have also used IRT to develop computerized adaptive testing (CAT) methods for efficiently evaluating latent traits \parencite{van2010}. The most prominent CAT method is the maximum information (item selection) method \parencite{lord1980}, which selects the item to maximize Fisher information and estimates the latent trait using the maximum likelihood estimator (MLE). The maximum information method has been widely used in fields such as education and clinical assessment \parencite{thomas2019} and has been shown that this procedure is asymptotically normal and efficient, making the maximum information method optimal \parencite{chang2009} .

However, the efficiency of the maximum information method is not guaranteed in the finite sample. A well-known problem is that, at the start of the CAT, the large bias of the MLE caused by the limited observations leads to information loss and compromises the efficiency of the maximum information method \parencite{Chang1999}. Another challenge is that the information maximization criterion cannot assign an item at the start of the test due to the cold start problem. Failures of assignment can also occur in the first few trials if the responses are all-correct or all-wrong \parencite{uniquenessMLE2024}, as the MLE is not well-defined in such cases. 
For these reasons, the maximum information method might not be ideal in a test with limited trials.

Bayesian approaches can circumvent the limitations mentioned above by incorporating prior knowledge of the distribution of abilities. For example, Bayesian approaches could avoid the cold start problem by using prior knowledge of the ability distribution. Moreover, the Bayes estimator can be seen as a balance between the prior and the responses, resulting in well-defined estimates for all possible outcomes. The most famous example is the weighted likelihood estimator. \textcite{magis2017} proved that the weighted likelihood estimator, as an estimator under Jeffery's prior distribution \parencite{Magis2012}, is finite under all possible responses. However, while Bayesian methods have already been developed in adaptive testing in various setups \parencite{vanderLinden1998, Ren2020}, a rigorous analysis of their statistical properties, especially their large sample guarantees, seems to be lacking.

This study aims to fill this gap by establishing a version of Bernstein-von Mises theorems for Bayesian adaptive tests \parencite{doob1949application,lecam1953,van2000asymptotic}.
\autoref{posterior converges}  and \autoref{Main Result 1} state our main results. \autoref{posterior converges} shows that, under regularity assumptions, the posterior distribution of the latent trait will concentrate around the truth at $\sqrt{n}-$ rate for any given sequence of item assignment.  
A direct application of the theorem shows that one can consistently estimate the latent trait by using either the posterior mean or the posterior median as the estimator. 
The arbitrariness in the selected items then implies
the consistency of Bayesian adaptive testing methods, 
which is presented in \autoref{Main Result 1}.
Moreover, the convergence is not limited to specific prior distributional assumptions as long as the prior includes the true parameter, which implies that the consistency of the estimator is robust to the mis-specification of the prior.

Although the main focus of this paper is to study the large-sample properties of general Bayesian testing procedures, we also propose a novel item selection procedure that optimizes the performance of the ability estimator. 
We first provide a rigorous foundation to define optimality by framing the item selection problem in the framework of Bayesian decision theory, which is then used to develop a novel item selection procedure. 
Unlike most existing adaptive methods that rely mostly on heuristics of optimality, our method is designed to optimize the performance of the ability estimator from the decision-theoretical perspective.

We also conducted Monte Carlo simulations to examine the statistical properties of our item selection method in the finite sample. Similar to the finding in \textcite{vanderLinden1998}, simulation results show that, compared to the classical method, estimators from the Bayesian procedure are more efficient in estimating latent traits, especially when the number of trials is limited. We also assessed our estimator's performance with different true latent traits and found that the Bayesian adaptive test is generally more efficient, except for extreme values of the true latent trait.

\begin{comment}
	Owen
	van der linden 的文文章中提出了根據maximaize infomation 或是 斯minimize variance 的方法，並針對大樣本性質進行論述，但該篇文章中僅是以貝氏方法大樣本下會收斂作為證明依據，然而這樣的證明並不嚴謹，因為在適性測驗中，每一回合回所面對的問題都與前一回合的回答有關，因此並非傳統問題所處理的iid 的結構，因此無法以簡單的方式論證該方法的收斂性。而在Owen 的研究中提出了minimize posterior variance 的方法和normal approximation method，然而在大樣本的證明也僅侷限於normal apprixmation method，並無法用於其他方法收斂性的證明。
\end{comment}
\begin{comment}
	1. Bayesian in IRT
	2. Bayesian IRT CAT
	3. Rigorous theoretical development
	4. gap in the Theorem and framework
	Q: 是否要 mention 我們的方法和linden 的方法會得到一樣的結果?
	
\end{comment}
\subsection{Literature Review}
In previous research, many scholars have used Bayesian estimation and inference to develop tests \parencite{Sijtsma2006}. The key feature of the Bayesian methodology is that it includes additional information in the process of estimation and inference, making Bayesian IRT appealing to researchers \parencite{fox2010bayesian}. %[Bayesian IRT] (why Bayesian?) 
There are many studies in the literature on the application of the Bayesian approach in CAT. For example, \textcite{owen1975} proposed the adaptive testing methods within a Bayesian framework through the normal approximation to the posterior distribution. \textcite{vanderLinden1998} also proposed an algorithm that assigns items by maximizing the expected information or minimizing the expected posterior variance. In this paper, we also propose a novel item selection method based on a decision theoretical framework.

While Bayesian methods have been widely used in adaptive testing, to the best of our knowledge, only a few studies have proven the mathematical properties of these methods while proposing them. For example, \textcite{Chang1993} only proved that under non-adaptive methods, the estimator developed under the assumption of a normal distribution prior is asymptotically normally distributed. As for \textcite{owen1975}, he only proved the consistency of Bayesian estimator  under the adaptive testing method he proposed. As for \textcite{vanderLinden1998}, in the section about the asymptotic property, he only mentioned that the convergence of the Bayesian estimator is a well-known fact. While the convergence is straightforward under the independence and identical distribution assumption, it might be violated in CAT. In the context of CAT, the distribution of the response in each trial depends on the previous responses, which makes the responses dependent and non-identical. Our main results show that, even under non-i.i.d. data generating process, we can still establish the asymptotic properties of Bayesian estimators under general item selection criteria.

The rest of the paper is organized as follows. In Section 2, we briefly introduce the IRT model and discuss some existing related adaptive methods. Next, we will define the tokens in the Bayesian decision theory and propose a new adaptive testing method in Section 3. In Section 4, we will prove the consistency of a general class of CAT methods under the IRT model. Finally, in Section 5, we will conduct Monte Carlo simulations and provide concluding remarks in the final section.

\section{Model and existing adaptive methods}

\subsection{Model Setup}
Consider the task of assessing a test-taker's ability on a given subject.
Let $X$ be the test-taker response on an item:
\begin{equation*}
	X=\begin{cases}
		1& \text{if the test-taker is correct on the item; }\;\;  \\
		0&	\text{otherwise.}
	\end{cases}
\end{equation*}
The probability of observing a correct response depends on the test-takers ability $\theta$ and item parameters $ \bm{\lambda}$. A classical model within the IRT framework is the Rasch model, which imposes that
\begin{equation}\label{Rasch}
	P(X=1\given  \theta, b)= \frac{1}{1+\exp(-(\theta-b))},
\end{equation}
where $\bm{\lambda}=b$ represent the item difficulty.

The conditional independence assumption implies that, given the difficulty parameter $\bm{b}_J=(b_1,...,b_J)$ and the subject's ability $\theta$, the likelihood of observing a sequence of responses $\bm{X}_J=(X_1,...,X_J)$  is:
\begin{equation*}
	L(\theta|\bm{b}_J,\bm{X}_J) = P(\bm{X}_J|\theta,\bm{b}_J)=\prod_{j=1}^J  P(X_j=1\given  {\theta},  b_j).
\end{equation*}
In the context of CAT, the parameters of items in the item bank are estimated beforehand so $\bm{b}_J=(b_1,...,b_J)$ can be treated as known. With the likelihood function, we can estimate the subject's ability using the maximum likelihood (ML) estimator:\footnote{In our simulation study, we use the constrained joint maximum likelihood estimation (CJMLE, \cite{Chen2018}). A benefit of the method is that it avoids the singularity of the MLE.} 
\begin{equation*}
	\hat{\theta}_{ML}=\arg \max_{\theta \in \Theta} L(\theta|\bm{b}_J,\bm{X}_J),
\end{equation*}
where $\Theta$ is the parameter space. 

The purpose of the CAT is to assign a sequence of items based on previous responses to estimate the test-taker's abilities efficiently. For example, the maximum  (Fisher) information method  \parencite{lord1980} assigns the next item in a way that the predicted probability of observing a correct response is exactly one-half. This method, which we will discuss further in the next subsection, is widely popular and extensively studied in the literature.

\subsection{The maximum information item selection method}

Under the Rasch model, given $J$ items, the (estimated) Fisher information is
\begin{equation*}
	I(\hat{\theta}\given \bm{b}_J,\bm{X}_J)=\sum_{j=1}^J P(X_j=1\given  \hat{\theta},  b_j)\;(1-P(X_j=1\given  \hat{\theta}, b_j))\;.
\end{equation*}

The maximum information method suggests that the next item should be chosen to maximize the above equation. Specifically, one should choose the item with difficulty $b^*$ such that
\begin{equation}\label{Information}
	b^*=\arg\max_{k \in S_{J+1}} I(\hat{\theta}\given \bm{b}_J,b_k)\;,
\end{equation}
where $S_{J+1}$ is the set of item indices remaining in the item bank in the $J+1$-th trial. The item that satisfies \eqref{Information} is the item with difficulty $b=\hat{\theta}_J$ \parencite{chang2009}, where $\hat{\theta}_J$ is the MLE based on the test-taker's responses to previous $J$ items. 
\textcite{chang2009} demonstrated that the maximum likelihood estimator (MLE) combined with the maximium information criteria is consistent and asymptotically normal under the Rasch model. They referred to this method as the ``optimal decision rule'' because its asymptotic variance achieves the minimum possible value under certain conditions. However, as mentioned in the introduction, this asymptotically optimal procedure can have sub-optimal finite-sample performance and can even be problematic in tests with only a few trials. 

\subsection{Bayesian estimators and item selection criteria 
	%\parencite{vanderLinden1998}
}
One approach to improve the finite-sample performance is to incorporate researchers' prior knowledge of the latent trait. Let $\pi$ be the prior distribution of the latent trait on the parameter space $\Theta$. By the Bayes' theorem, upon observing the responses $\bm{X}_J=(X_1,...,X_J)$, the posterior distribution of the latent trait is given by:
\begin{equation*}
	\pi_{\theta|\bm{x}_J,\bm{b}_J}(\theta)=\frac{ P(\bm{X}_J=\bm{x}_J|\theta,\bm{b}_J)\pi(\theta)}{\int P(\bm{X}_J=\bm{x}_J|\theta,\bm{b}_J)\pi(\theta) \text{d}\theta}.
\end{equation*}
Given the posterior distribution, there are several ways to obtain the point estimator of $\theta$ given $\bm{x}_J$ and $\bm{b}_J$. For example, researchers can use the posterior mean or posterior mode as an estimator:
\begin{align*}
	\hat{\theta}_{mean} &= \int_{\Theta} \theta  \pi_{\theta|\bm{x}_J,\bm{b}_J}(\theta) \mathrm{d}\theta,\\
	\hat{\theta}_{mode} &=  \arg \max_{\theta \in \Theta}    \pi_{\theta|\bm{x}_J,\bm{b}_J}(\theta)
\end{align*}
if it is desired to minimize the mean squared error (MSE) or the mean absolute error. The main result of this paper shows that the posterior distribution concentrates around the true ability whenever the item selection procedure satisfies certain regularity assumptions. As a consequence, both the posterior mean and mode will converge to the true value as the sample size goes to infinity. Our result gives large-sample, frequentist-style guarantees for Bayesian estimators, and it applies to a wide class of item selection procedures.\footnote{The item selection criterion itself needs not be Bayesian.}

Prior information can also be incorporated in the item selection criterion. For example, \textcite{vanderLinden1998} follows the idea of the maximum information method and extends it by averaging the Fisher information over the prior distribution of the latent trait. Specifically, the next item is chosen to maximize the posterior-weighted information
\begin{equation*}
	b^*=\arg \max_{b \in S_{J+1} } \int I({\theta}\given b) \pi_{\theta|\bm{x}_J,\bm{b}_J}(\theta) d \theta,
\end{equation*}
where $\pi_{\theta|\bm{x}_J,\bm{b}_J}(\theta)$ is the posterior distribution of $\theta$ given previous response $\bm{x}_J$ and $\bm{b}_J$. In \textcite{vanderLinden1998}, several other criterion for optimality are also proposed, including the maximum expected information criteria, the minimum expected posterior variance criteria, and the maximum expected posterior-weighted information criteria. Notice that, while these procedures are designed to optimize certain objective function, they generally lack direct connection to the performance of the final ability estimator. We aim to bridge this gap by analyzing the adaptive problem within the framework of Bayesian decision theory and by proposing a new item selection procedure that directly optimizes the accuracy of the resulting Bayes estimator.

\section{A Bayesian decision-theoretical approach to the Adaptive Problem}

We now formally state the adaptive testing problem in a decision-theoretical framework. Consider a test with $J$ items and define the estimator of the latent ability as $\hat{\theta}_J=f(\bm{B}_J,\bm{X}_J)$, where $\bm{B}_J=(B_1,...,B_J)$ is the sequence of the item difficulties and $\bm{X}_J=(X_1,...,X_J)$ is the response vector. In CAT, the decision problem is to choose the sequence of item difficulty $\bm{B}_J = (B_1, B_2, ..., B_J)$ and the estimator of the latent ability $\hat{\theta}_J$ to minimize the expected loss (also known as the risk) $E[L(\hat{\theta}, \theta)]$, where the expectation is taken over the distribution of the true ability \( \theta \) and the response vector \( \bm{X}_J \). The loss function \( L(\hat{\theta}, \theta) \) measures the discrepancy between the estimated ability \( \hat{\theta} \) and the true latent ability \( \theta \). Commonly used loss functions include the squared error loss, \( (\hat{\theta} - \theta)^2 \), which penalizes larger deviations more heavily, and the absolute error loss, \( |\hat{\theta} - \theta| \), which treats all deviations proportionally.

%We aim to find the estimator which on average minimizes the loss across the distribution of $\theta$ and $X$.

A decision rule is a function that maps the past information (the responses and difficulties of the items assigned) into the decision. An important point is to use all the information available. Therefore, instead of assigning all $B_j$'s before the test starts, we choose each $B_j$ by taking into account the all the information available at the time of the decision being made. That is, to choose $B_j$, the decision maker can make use of the following information: (i) $\pi$, the prior belief on the latent ability $\theta$, (ii) the agent's previous responses $\bm{X}_{j-1}$, and (iii) the item difficulties $\bm{B}_{j-1}$. Therefore, for each $1 < j \leq J$, $B_j$ depends on previous responses and the prior distribution $\pi$, and so we write $B_j = b_j(\bm{X}_{j-1},\bm{B}_{j-1}; \pi) = b_j(\bm{X}_{j - 1},\bm{B}_{j-1})$. For $j=1$, since we have not yet seen any response from the agent, all the information we have is the prior distribution. Therefore, $B_1$, the difficulty of the first problem $B_1$ has no randomness, and its value of $B_1 = b_1$ only depends on $\pi$.

Our objective is to find the rule $\delta_J = (\hat{\theta}_J, \{b_j\}_{j = 1}^J)$ that minimizes expected loss.

Given prior $\pi$, for any the Bayes risk $r(\pi, \delta_J)$ of a decision rule $\delta_J$ is defined as
\begin{align*}
	r(\pi, \delta_J) = \int \sum_{\bm{x}_J}L(\hat{\theta}_J(\bm{x}_J, \bm{b}_J), \theta) \P(\bm{X}_J = \bm{x}_J, \bm{B}_J = \bm{b}_J \given \theta, \delta_J) d\pi(\theta),
\end{align*}
where 
\begin{align*}
	&\quad \P(\bm{X}_J = \bm{x}_J, \bm{B}_J = \bm{b}_J \given \theta, \delta_J) \\
	&= p(\theta, b_1)^{x_1}(1 - p(\theta, b_1))^{1 - x_1} 
	\prod_{j = 2}^J p(\theta, b_j(\bm{x_{j - 1}}))^{x_j} (1 - p(\theta, b_j(\bm{x_{j - 1}})))^{1 - x_j}.
\end{align*}
The choice of $\hat{\theta}_J$ has been well established in the Bayesian literature. For example, when the loss is squared loss,  $\hat{\theta}_J$ is the posterior mean; 
and, when the loss is absolute loss, $\hat{\theta}_J$ is the posterior median. The remaining question is how to choose the item difficulties $b_j$'s. If $J = 1$, then we are only choosing $b_1$, and the optimal $b_1$
depends only on the prior. However, when $J > 1$, deriving the optimal rule requires solving a dynamic problem which generally involves complex calculation, making it impractical.

We propose a iterative method to solve to the decision problem. After observing the agent's responses to the first $j-1$ items, we choose the difficulty $b_{j}$ of the next item to be the one that minimizes the expected loss after we observe the agent's first $j$ responses. That is, if the posterior distribution after observing the agent's responses to the first $j$ items is $\tilde{\pi}$, we iteratively find the optimal item by solving 
\[
\min_{\delta} r(\tilde{\pi}, \delta).
\]
In below, we lay out the iterative procedure for selecting item difficulties and estimating the agent's latent ability:
\begin{enumerate}
	\item Find $(\hat{\theta}, b_1) = \arg\min_{\delta}{r(\pi,\delta)}$. Assign the first item difficulty to be $b_1$.
	\item For $j \geq 2$,
	\begin{enumerate}[(i)]
		\item Upon observing the agent's previous response $x_{j - 1}$ to the item with difficulty $b_{j - 1}$, 
		update the distribution $\pi$ by calculating the posterior distribution $\tilde{\pi}$.
		\item Suppose $(\hat{\theta}^*, b^*) = \arg \min_{\delta} r(\tilde{\pi}, \delta)$. 
		Choose $b_{j} = b^*$ and replace $\pi$ by $\tilde{\pi}$.
	\end{enumerate}
	\item Repeat step 2 until $j=J$. After observing the agent's response to the item with difficulty $b_J$,
	calculate the Bayesian estimator $\hat{\theta}_J$ using the posterior distribution.
\end{enumerate}

Note that, unlike the existing item selection methods such as the ones we introduced in the previous section, the objective function we consider here has a direct connection to the performance (i.e., the Bayes risk) of the ability estimator. By explicitly linking item selection to the performance of the consequential ability estimator, this approach aligns the testing process with the ultimate goal of producing accurate and reliable estimates of the latent trait. Moreover, the iterative nature of the method allows for dynamic adjustments based on observed responses, making it both flexible and computationally feasible in practice.

\section{Consistency of Bayesian Estimator under Adaptive Testing}
In this section, we give sufficient conditions that an adaptive item selection criterion will generate a consistent Bayesian ability estimator; that is, as the number of questions goes to infinity, the estimator of a subject's ability converges in probability to its value.

A main challenge in proving the convergence is the dependence between a subject's responses. 
Because of the sequential nature of the item selection criterion, the subject's responses to each item are neither independent nor identically distributed (non-i.i.d.). 
First, since the difficulty of the next item depends on previous responses, the distribution of the next response depends on previous responses.
Second, since the difficulty of the item may change, the distribution of responses may also change over time. 
The sequential dependence renders classical theorems based on independence and identical distribution inapplicable, necessitating more nuanced techniques to establish the desired convergence.

%Notice that, once we condition on the difficulty of the items the responses become independent (though still not identically distributed).
To address these challenges, we apply a powerful theorem (\autoref{GV}) in \textcite{GV} that establishes the convergence for posterior distribution with non-i.i.d. observations.
We provide a general result (\autoref{posterior converges}) that ensures the posterior distribution will concentrate around the true ability given a sequence of item difficulties $\{b_j\}$, and, as a consequence, both the posterior mean and posterior median converge to the true ability (\autoref{mean consistency}).
With this established, if we use either the posterior mean or the posterior median in CAT to estimate the subject's ability, the ability estimator converges to the true ability (\autoref{Main Result 1}).
The two assumptions needed are listed below.
\begin{assumption} \label{bounded parameter space}
	The parameter space $\Theta \subset \R$ is bounded, and the prior distribution $\Pi$ admits a p.d.f. $\pi$ that is strictly positive and bounded above on $\Theta$.
\end{assumption}

\begin{assumption} \label{bounded problem set}
	The adaptive item selection method chooses difficulties of the items from a bounded set $\mathcal{B} \subset \R$.
\end{assumption}

For the posterior mode, the following assumption is needed to guarantee its existence and uniqueness. Under this assumption, we can show that the posterior mode also converges to the true ability for a given sequence of item difficulty. 
This convergence can then be used to show that the posterior mode is a consistent estimator of the latent ability when combined with an adaptive item selection procedure (\autoref{Main Result 2}).

\begin{assumption} \label{logconcave}
	The parameter space $\Theta$ is a closed interval, and the prior distribution $\Pi$ admits a strictly positive logconcave p.d.f. $\pi$ on $\Theta$.
\end{assumption}
Classic examples of logconcave distributions defined on a closed interval are truncated normal distribution and uniform distribution.
Before stating and proving our results, we first define some notations.
\begin{itemize}
	\item We use capital letters to represent random variables/sequences while smaller letters to represent 
	realizations. For example, $\bm{X}_J$ means a random sequence of 
	responses, while $\bm{x}_j$ means a realized sequence of responses.
	\item $G(t) \coloneq \exp(t) / (1 + \exp(t))$ and $\overline{G}(t) = 1 - G(t)$.
	\item $\Pi_J( \cdot \given \bm{x}_J, \bm{b}_J)$ denotes the posterior distribution of the subject's ability $\theta$ over $\Theta$ upon observing the agent's first $J$ responses $\bm{x}_J = \{x_1,...,x_J\}$ with item difficulties $\bm{b}_J = \{b_1,...,b_J\}$.
	\item $\hat{\theta}^{mean}_J(\bm{x}_J \given \bm{b}_J)$, $\hat{\theta}^{med}_J(\bm{x}_J \given \bm{b}_J)$ 
	and $\hat{\theta}^{mode}_J(\bm{x}_J \given \bm{b}_J)$ denote the posterior mean, posterior median and posterior 
	mode upon observing the agent's first $J$ responses $\bm{x}_J = \{x_1,...,x_J\}$ with item difficulties $\bm{b}_J = \{b_1,...,b_J\}
	$. 
	\item $\hat{\theta}^{adp}_J$ denote the ability estimator of the agent's ability provided by an adaptive testing method after observe first $J$ items and observing the agent's responses to those items.
	If we use posterior mean to estimate the subject's ability in CAT, then $\hat{\theta}^{adp}_J = \hat{\theta}^{mean}_J(\bm{X}_J \given \bm{B}_J)$, 
	where the distribution of $\bm{X}_J$ and $\bm{B}_J$ depends on the 
	item selection criterion and the subject's true ability.
\end{itemize}
\autoref{posterior converges} shows that, given a sequence of items, the posterior distribution will concentrate on the true latent ability. \autoref{mean consistency} further establishes that the posterior mean and median also converge to the true value.
\begin{theorem} \label{posterior converges}
	Fix a bounded sequence of difficulties $\bm{b}_\infty = \{b_j\}_{j = 1}^\infty$, and suppose
	\autoref{bounded parameter space} holds.
	Then for any true ability $\theta_0 \in \Theta$, the posterior distribution converges to the true parameter $\theta_0$
	at rate $\sqrt{J}$, that is, for any $M_J \to \infty$,
	\begin{equation*}
		\int_{[0,1]^\infty} \Pi_J(|\theta - \theta_0| > \sqrt{J} M_J \given \bm{x}_J, \bm{b}_J) \dd \P_0(\bm{x}_\infty) \longrightarrow 0
	\end{equation*}
	as $J \to \infty$,
	where $\{0, 1\}^\infty$ is the sample space of all possible sequences of responses $\bm{x}_\infty$, 
	and $\P_0$ is the probability measure over $\Theta$ induced by the subject's true ability $\theta_0$
	and the fixed sequence of item difficulties $\bm{b}_\infty$.
	It follows that 
	\begin{equation*}
		\Pi_J(|\theta - \theta_0| > \sqrt{J} M_J \given \bm{X}_J, \bm{b}_J) \pto 0.
	\end{equation*}
	as $J \to \infty$.
\end{theorem}
\begin{proof}
	We check that the condition of Theorem 10, \cite{GV} holds and apply the theorem. 
	A detailed verification of the theorem's condition is presented in \autoref{proofs}.
\end{proof}

\begin{corollary} \label{mean consistency}
	Suppose the sequence of difficulties $\bm{b}_\infty = \{b_k\}$ is fixed and bounded, and 
	\autoref{bounded parameter space} holds.
	Then given any true ability $\theta_0 \in \Theta$,
	$\hat{\theta}^{mean}_J(\bm{X}_J \given \bm{b}_J)$ and the posterior median 
	$\hat{\theta}^{med}_J(\bm{X}_J \given \bm{b}_J)$ upon observing the subject's responses of the first $J$ items $\bm{X}_J$ converges to $\theta_0$ in probability.
\end{corollary}
\begin{proof}
	Let $D = \max\{|\theta_0 - \inf \Theta|, |\theta_0 - \max \Theta|\}$.
	Let $\epsilon > 0$ be small, and consider $\eta = \frac{\epsilon}{2}$ and $p = \frac{D - \epsilon}{D - \frac{\epsilon}{2}}$.
	The event $\{\,|\hat{\theta}^{mean}_J - \theta_0| \leq \epsilon \,\}$ contains the event 
	$\{\, \Pi(|\theta - \theta_0| \leq \eta \given \bm{X}_J) \geq p \,\}$ because when the latter happens,
	\begin{align*}
		|\hat{\theta}^{mean}_J(\bm{X}_J) - \theta_0| &\leq \eta \Pi(|\theta - \theta_0| < \eta \given \bm{X}_J) + D(1 - \Pi(|\theta - \theta_0| < \eta \given \bm{X}_J) ) \\
		&\leq \eta p + D(1 - p) \leq \epsilon.
	\end{align*}
	By \autoref{posterior converges}, for any $\eta > 0$ and $p < 1$, 
	the probability of the event $\{ \Pi_J(|\theta - \theta_0| \leq \eta \given \bm{X}_J, \bm{b}_J) \geq p \}$  converges to $1$,
	and therefore, the probability of the event $\{\,|\hat{\theta}^{mean}_J - \theta_0| \leq \epsilon \,\}$
	converges to 1. We conclude that $\hat{\theta}^{mean}_J$ converges to $\theta_0$ in probability.
	Similarly, one can show that for any small $\epsilon > 0$, there exists a small $\eta > 0$ and a large $p < 1$ 
	such that the event $\{\, |\hat{\theta}^{med}_J - \theta_0| < \epsilon \,\}$ contains 
	$\{\, \Pi(|\theta - \theta_0| < \eta \given \bm{X}_J) \geq p \,\}$, and thus obtain the fact that $\hat{\theta}^{med}_J$ converges to $\theta_0$
	in probability.
\end{proof}
Both \autoref{posterior converges} and \autoref{mean consistency} consider the case when the item difficulty is exogenously given. However, in adaptive item selection methods, the item difficulties are generally functions of past observations, and the randomness and dependence in the item selected must be accounted. \autoref{Main Result 1} extends the convergence result to the case when item difficulties are endogenously chosen based on an item selection method. It shows that, Bayesian estimators, are consistent estimators for the latent ability when combined with adaptive item selection criteria.
\begin{theorem} \label{Main Result 1}
	Suppose \autoref{bounded parameter space} and \autoref{bounded problem set} hold.
	If we use the posterior mean or the posterior median combined with the adaptive item selection method,
	then the estimator generated from the method $\hat{\theta}^{adp}_J$ converges to $\theta_0$ in probability.
\end{theorem}
\begin{proof}
	Let $S= \{0, 1\}^\infty \times \mathcal{B}^\infty$ be the sample space of all possible 
	$\bm{x}_\infty$ and $\bm{b}_\infty$.
	Let $\P_0$ be the probability measure over $S$ induced by the subject's true ability $\theta_0$ and the adaptive method, 
	$\P_0^b$ be the marginal probability measure of $\bm{b}_\infty$ over $\Theta^\infty$,
	and $\P_0^x(\cdot \given \bm{b}_\infty)$ be the conditional probability measure of $\bm{x}_\infty$
	over $[0, 1]^\infty$ given any sequence of difficulty $\bm{b}_\infty$.
	Given any $\epsilon > 0$, 
	\begin{align*}
		\P_0\{ |\hat{\theta}^{adp}_J - \theta_0| > \epsilon\}
		&= \int_{S} \indicator{|\hat{\theta}^{adp}_J - \theta_0| > \epsilon} \dd \P_0 \\
		&= \int_{\mathcal{B}^\infty} \int_{\{0, 1\}^\infty} \indicator{|\hat{\theta}^{adp}_J - \theta_0| > \epsilon} \dd \P_0^x(\bm{x}_\infty \given \bm{b}_\infty) \dd \P_0^b(\bm{b}_\infty) \\
		&= \int_{\mathcal{B}^\infty} \P\{|\hat{\theta}^{mean}_J(\bm{X}_J \given \bm{b}_J) - \theta_0| > \epsilon \} \dd \P_0^b(\bm{b}_\infty)
	\end{align*}
	By \autoref{mean consistency}, for any $\bm{b}_\infty \in \mathcal{B}^\infty$,
	\[ \P_0\{|\hat{\theta}^{mean}_J(\bm{X}_J \given \bm{b}_J) - \theta_0| > \epsilon \} \longrightarrow 0\]
	as $J \to \infty$. Hence, by the Bounded Convergence Theorem, 
	\[ \P_0\{ |\hat{\theta}^{adp}_J - \theta_0| > \epsilon\} \longrightarrow 0\]
	as $J \to \infty$.
\end{proof}
Similar results can also be established for the posterior mode, assuming it is well-defined and unique. \autoref{lemma1} provides a sufficient condition for the posterior mode to be both well-defined and unique. \autoref{mode consistency} establishes the convergence of the posterior mode when conditioning on item difficulty, and \autoref{Main Result 2} establishes convergence without conditioning on item difficulty.
\begin{lemma}\label{lemma1}
	Suppose \autoref{logconcave} holds. 
	Fixing any sequence of difficulties $\bm{b}_\infty = \{b_j\}$ and realized responses $\bm{x}_\infty = \{x_j\}$, 
	the posterior distribution $\pi_n(\theta \given \bm{x}_J, \bm{b}_J)$ is log-concave and thus unimodal for all $n$.
	It follows that the posterior mode 
	\[ \hat{\theta}^{mode}_J(\bm{x}_J \given \bm{b}_J) = \arg \max_{\theta \in \Theta}\pi_n(\theta \given \bm{x}_J, \bm{b}_J) \]
	is well-defined and unique. (The mode may be on the boundary of $\Theta$.)
\end{lemma}
\begin{proof}
	Prove by induction. Suppose $\pi_{n-1}(\theta) = \pi_{n-1}(\theta \given \bm{x}_{n-1}, \bm{b}_{n-1})$
	is log-concave.
	Then 
	\begin{align*}
		\frac{d}{d\theta}\pi_n(\theta \given \bm{x}_J, \bm{b}_J) = 
		\begin{cases}
			\frac{d}{d\theta} \log \pi_{n-1}(\theta) + \frac{d}{d\theta} \log G(\theta - b_n) & \text{ if } x_n = 1 \\
			\frac{d}{d\theta} \log \pi_{n-1}(\theta) + \frac{d}{d\theta} \log \overline{G}(\theta - b_n) & \text{ if } x_n = 0.
		\end{cases}
	\end{align*}
	Since $G(\theta - b)$ and $\overline{G}(\theta - b)$ are log-concave, $\pi_n(\theta \given \bm{x}_J, \bm{b}_J)$
	must be log-concave and thus unimodal. It then follows that $\hat{\theta}^{mode}_J(\bm{x}_J \given \bm{b}_J)$ is unique for all $n$.
	$\hat{\theta}^{mode}_J(\bm{x}_J \given \bm{b}_J)$ exists because $\Theta$ is a closed interval.
\end{proof}

\begin{corollary} \label{mode consistency}
	Suppose \autoref{logconcave} holds, and fix any bounded sequence of difficulties $\bm{b}_\infty = \{b_k\}$.
	Then the posterior mode $\hat{\theta}^{mode}_J = \hat{\theta}^{mode}_J(\bm{X}_J \given \bm{b}_J)$ of observing the first $J$ responses $\bm{X}_J$ converges in probability to the true ability $\theta_0$. 
\end{corollary}

\begin{proof}
	By \autoref{posterior converges}, for any $\eta > 0$ and $p < 1$, 
	the probability of the event $\{ \Pi_J(|\theta - \theta_0| \leq \eta \given \bm{X}_J, \bm{b}_J) \geq p \}$  converges to $1$.
	Let $\epsilon > 0$ be small and simply set $\eta = \frac{\epsilon}{4}$ and $p = \frac{2}{3}$. 
	Then the event $\{|\hat{\theta}^{mode}_J - \theta_0| \leq \epsilon \}$ contains the event 
	$\{ \Pi_J(|\theta - \theta_0| \leq \eta \given \bm{X}_J, \bm{b}_J) \geq p \}$. 
	The statement follows from the observation that when $\epsilon > 0$ is small and $|\hat{\theta}^{mode}_J - \theta_0| > \epsilon $,
	\[ \Pi_J\left(|\theta - \theta_0| \leq \frac{1}{4}\epsilon \given[\Big] \bm{x}_J, \bm{b}_J\right) < \Pi_J\left( |\theta - \hat{\theta}^{mode}_J| < \frac{3}{4} \epsilon \given[\Big] \bm{x}_J, \bm{b}_J \right)\]
	for any $\bm{x}_J$, and thus the left-hand side must be smaller than $\frac{1}{2}$.
	Hence when $ \Pi_J(|\theta - \theta_0| \leq \eta \given \bm{X}_J, \bm{b}_J) \geq \frac{2}{3} $, 
	$|\hat{\theta}^{mode}_J - \theta_0| \leq \epsilon $.
	By the statement, we conclude that the probability of $\{|\hat{\theta}^{mode}_J - \theta_0| \leq \epsilon \}$ converges to 1, 
	and thus $\hat{\theta}^{mode}_J \pto \theta_0$.
\end{proof}

\begin{theorem} \label{Main Result 2}
	Suppose  \autoref{bounded problem set} and \autoref{logconcave} hold.
	If we use the posterior mode in the adaptive method,
	then the estimator generated from the method $\hat{\theta}^{adp}_J$ converges to $\theta_0$ in probability.
\end{theorem}
\begin{proof}
	With the previous corollary established, the proof is identical to that of \autoref{Main Result 1}.
\end{proof}

\section{Simulation}
In this section, we aimed to compare the performance of the estimators based on Bayes decision theory and the estimator based on the information criterion. To this end, we simulate the subjects' responses with different thetas under the assumption of the Rasch model. In this study, we conducted simulations under the assumption that the distribution of ability follows a normal distribution \parencite{Linden1997}. To assess the performance of different testing methods while considering the stochastic nature of ability, we systematically sampled quantiles from the normal distribution as values for $\theta$. We performed 100 simulation experiments for each of these theta values to evaluate the effectiveness of two methods under the Rasch model assumption.

Each simulation experiment aimed to mimic examinees' responses in 30 trials of Computerized Adaptive Testing (CAT), providing estimates for each trial. Mean Squared Error (MSE) was employed as the evaluation metric, measuring the performance of different methods under the Rasch model assumption. The MSE was tracked across trials to observe variations in performance for different methods.

Furthermore, to gain insights into the performance of different methods at specific ability levels, we examined the MSE for each method at different $\theta$ values when the number of trials is 30. All simulations were performed in R \parencite{R}.

\begin{figure}
	\centering
	\includegraphics[width=\textwidth]{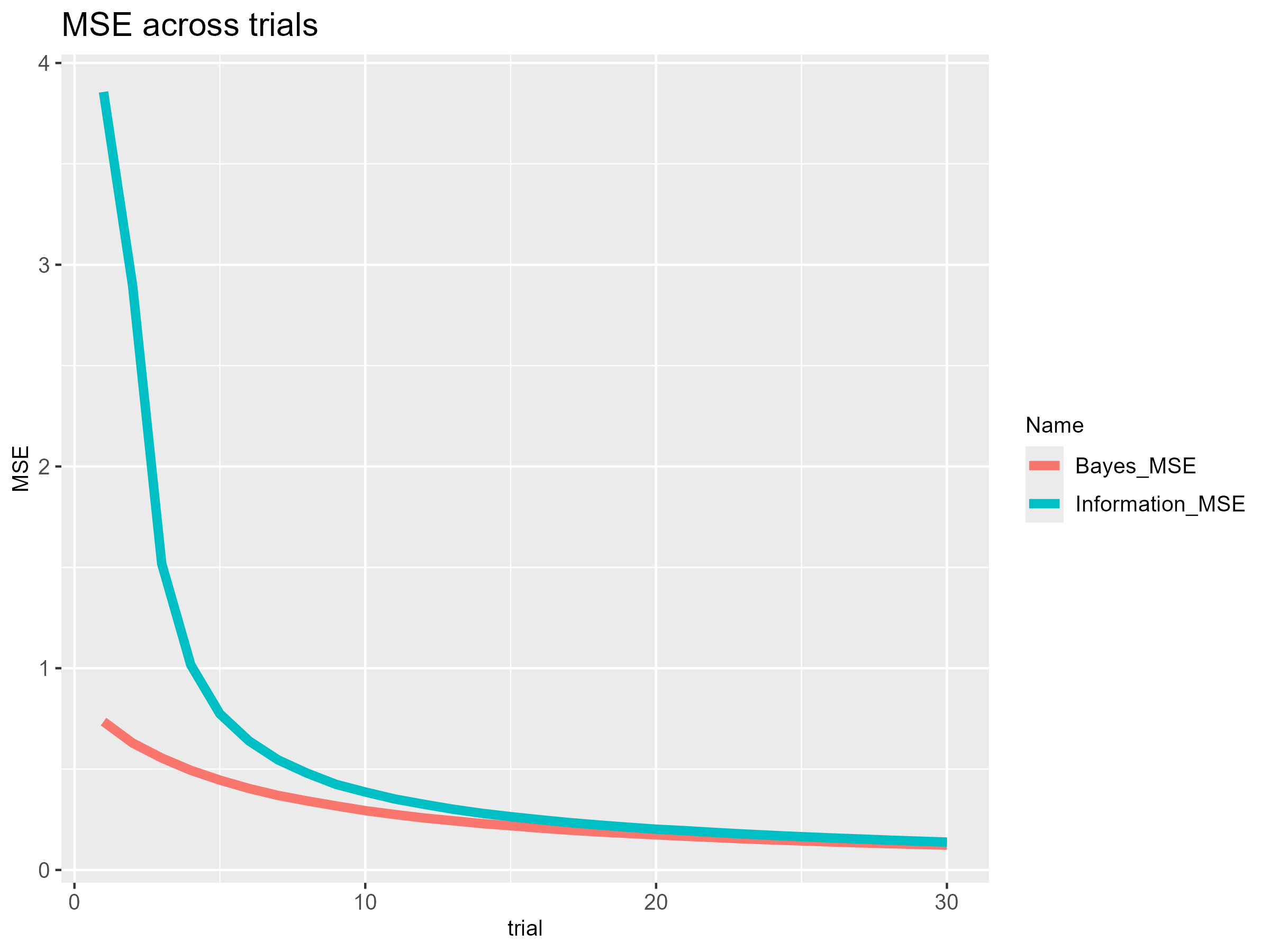}
	\caption{MSE of different methods across trials}\label{MSE_trial}
\end{figure}

\begin{figure}
	\centering
	\includegraphics[width=\textwidth]{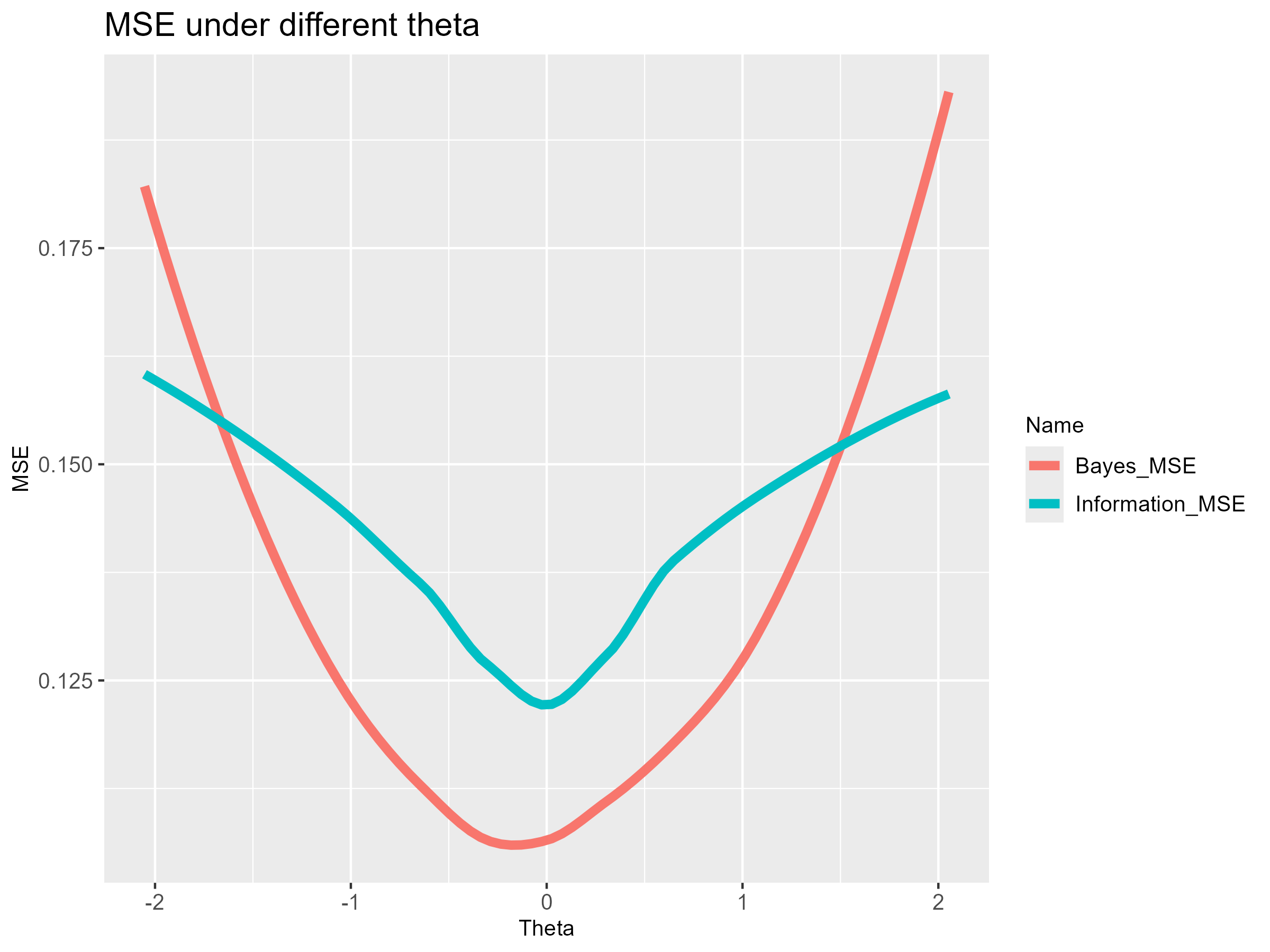}
	\caption{MSE of different methods across different $\theta$}\label{MSE_theta}
\end{figure}

\autoref{MSE_trial} presents the Mean Squared Error (MSE) of the Sequential method and the traditional Information Maximization method across different trials. We observe that while the MSE of both methods decreases as the number of trials increases, the rate of decrease in MSE for the sequential method is significantly better than that for the Information method, particularly when the number of rounds is less than 10, where the MSE of the former method already shows a notable decline. Furthermore, we find that the performance of the Sequential method at the tenth trial is already comparable to that of the Information method at the twentieth trial, supporting the superiority of the Sequential method in small-sample scenarios. \autoref{MSE_theta} displays the MSE of both methods at the thirtieth trial under different true values of $\theta$. From the figure, we can see that except under extreme values of $\theta$ where the Information method performs better than the sequential method, the Sequential method generally exhibits superior performance in most situations.

%\subsection{Pointwise performance of the Bayes estimator}

\section{Concluding Remarks}

This study provides a rigorous foundation for Bayesian adaptive testing under the Rasch model, bridging the gaps in the theoretical understanding of these methods. By proving the consistency of Bayesian estimators under adaptive item selection, we show that Bayes estimators converge to the true latent trait even under non-i.i.d. item assignment. Additionally, our Bayesian decision-theoretical approach introduces a novel item selection procedure that directly optimizes the performance of the ability estimator, aligning the adaptive testing process with the ultimate goal of producing accurate and reliable measurements.

The simulations results highlight the efficiency of the proposed Bayesian adaptive methods, particularly in small-sample scenarios where traditional maximum information approaches often struggle. These findings underscore the practical relevance of incorporating Bayesian principles into adaptive testing, offering a pathway for improved performance in empirical applications. Future research may extend this framework to more complex models or explore its integration with emerging computational techniques.
\printbibliography
%\bibliography{reference}
\appendix 
\section{Proofs} \label{proofs}

\begin{theorem}[Theorem 10, \cite{GV}] \label{GV}
	Let $X_1, ..., X_n$ be independent and non-identical observations on the sample space $\mathcal{X}$ following distributions $\P_{\theta, i}$, where $\Theta \subset \R^d$.
	Assume for each $\theta \in \Theta$ and 
	$i \in \mathbb{N}$, $\P_{\theta, i}$ is 
	dominated by a $\sigma$-finite measure $\mu$
	with density $p_{\theta, i}$.
	Let the true parameter $\theta_0$ be an interior point of $\Theta$. Assume that there exists constants $\alpha > 0$ and 
	$0 \leq c_i \leq C_i < \infty$ with, for every $\theta, \theta_1, \theta_2 \in \Theta$,
	\[ c = \liminf_{n \to \infty} \frac{1}{n} \sum_{i = 1}^n c_i > 0, \quad C = \limsup_{n \to \infty} \frac{1}{n} \sum_{i =1}^n C_i < \infty\]
	such that 
	\begin{align*}
		&\int_{\mathcal{X}} \log \left( \frac{p_{\theta_0, i}}{p_{\theta, i}} \right) \dd \P_{\theta_0, i} \leq C_i \norm{\theta - \theta_0}^{2\alpha} \\
		&\int_{\mathcal{X}} \log \left( \frac{p_{\theta_0, i}}{p_{\theta, i}}\right)^2 \dd \P_{\theta_0, i} \leq C_i \norm{\theta - \theta_0}^{2\alpha} \\
		&c_i \norm{\theta_1 - \theta_2}^{2 \alpha} \leq 
		h^2(p_{\theta_1, i}, p_{\theta_2, i}) \leq 
		C_i \norm{\theta_1 - \theta_2}^{2 \alpha}.
	\end{align*}
	where $h^2(p_{\theta_1, i}, p_{\theta_2, i})$ is the Hellinger 
	distance between $p_{\theta_1, i}$ and $p_{\theta_2, i}$,
	\begin{align*}
		h^2(p_{\theta_1, i}, p_{\theta_2, i}) = 
		\int_{\mathcal{X}} (\sqrt{p_{\theta_1, i}} - \sqrt{p_{\theta_2, i}})^2 
		\dd \mu.
	\end{align*}
	Assume that the prior measure $\Pi$ possesses a density $\pi$ which is 
	bounded away from zero in a neighborhood of $\theta_0$ and bounded above 
	on the entire parameter space.
	Then the posterior converges at the rate $n^{-1/(2\alpha)}$
	with respect to the Euclidean metric.
\end{theorem}

\begin{proof}[proof of \autoref{posterior converges}]
	{\footnotesize
		Define
		\[ G(t) = \frac{1}{1 + t}, \quad \overline{G}(t) = 1 - G(t).\]
		In the context of Rasch model, the sample space $\mathcal{X} = \{ 0, 1\}$ and we can set the 
		$\sigma$-finite measure $\mu$ to be $\mu(\{0\}) = \mu(\{1\}) = 1$.
		Then each response $X_j$ with item difficulty $b_j$ has density
		\[ p_{\theta, j}(1) = G(\theta - b_j), \quad p_{\theta, j}(0) = \overline{G}(\theta - b_j).\]
		Define for any $\theta_1, \theta_2 \in \Theta$ and $b \in \{ b_j \}_{j = 1}^\infty$,
		\begin{align*}
			V_{1b}(\theta_1 \from \theta_2) &= \left( \log \frac{G(\theta_1 - b)}{G(\theta_2 - b)} \right)G(\theta_1 - b)
			+ \left( \log\frac{\overline{G}(\theta_1 - b)}{\overline{G}(\theta_2 - b)} \right) \overline{G}(\theta_1 - b) \\
			V_{2b}(\theta_1 \from \theta_2) &= \left( \log \frac{G(\theta_1 - b)}{G(\theta_2 - b)} \right)^2G(\theta_1 - b)
			+ \left( \log\frac{\overline{G}(\theta_1 - b)}{\overline{G}(\theta_2 - b)} \right)^2 \overline{G}(\theta_1 - b) \\
			h_b^2(\theta_1, \theta_2) &= \left(\sqrt{G(\theta_1 - b)} - \sqrt{G(\theta_2 - b)}\right)^2 + \left(\sqrt{\overline{G}(\theta_1 - b)} - \sqrt{\overline{G}(\theta_2 - b)}\right)^2
		\end{align*} 
		To check the conditions of Theorem 10, \cite{GV}, it suffices to prove that there exists $0< c < C < \infty$ such that 
		for all $\theta_1, \theta_2 \in \Theta$ and $b \in \{b_j\}$,
		\begin{enumerate}
			\item $V_{1b}(\theta_1 \from \theta_2) \leq C | \theta_1 - \theta_2 |^2$
			\item $V_{2b}(\theta_1 \from \theta_2) \leq C | \theta_1 - \theta_2 |^2$
			\item $c|\theta_1 - \theta_2|^2 \leq h_b^2(\theta_1, \theta_2) \leq C|\theta_1 - \theta_2|^2 $
		\end{enumerate}
		We first prove that $C$ can be chosen to be 4.
		By computation, we obtain \[ \vpd{}{\theta_2} V_{1b}(\theta_1 \from \theta_2) = G(\theta_2 - b) - G(\theta_1 - b).\]
		Note that $V_{1b}(\theta_1 \from \theta_1) = 0$ and $V_{2b}(\theta_1 \from \theta_1) = 0$. 
		Using Mean Value Theorem iteratively, for some $\theta^*$ between $\theta_1$ and $\theta_2$,
		$\theta^{**}$ between $\theta_1$ and $\theta^*$,
		\begin{align*}
			V_{1b}(\theta_1 \from \theta_2) &= V_{1b}(\theta_1 \from \theta_2) - V_{1b}(\theta_1 \from \theta_1) \\
			&= (G(\theta^* - b) - G(\theta_1 -b))(\theta_2 - \theta_1) \\
			&= G(\theta^{**} - b)\overline{G}(\theta^{**} - b)(\theta^* - \theta_1)(\theta_2 - \theta_1). \\
			&\leq \frac{1}{4}|\theta_2 - \theta_1|^2.
		\end{align*}
		Indeed, we also know that $V_{1b}(\theta_1 \from \theta_2) \leq |\theta_1 - \theta_2|$
		from the second equation.
		By computation, we obtain
		\[ \vpd{}{\theta_2}V_{2b}(\theta_1 \from \theta_2) = 2V_{1b}(\theta_1 \from \theta_2)G(\theta_2 - b) 
		- 2\left(\log\frac{G(\theta_1 - b)}{G(\theta_2 - b)} \right)G(\theta_1 - b). \]
		Define 
		\[ h_b^2(x \from y) = \left(\log\frac{G(x - b)}{G(y - b)} \right)G(x - b) .\] for any $x, y, b \in \Theta$.
		Then $h_b^2(x \from x) = 0$ and $\frac{\partial}{\partial y}h_b^2(x \from y) = -G(x - b) \overline{G}(y-b)$.
		By Mean value Theorem, for some $\theta^*$ between $\theta_1$ and $\theta_2$,
		and $\theta^{**}$ between $\theta_1$ and $\theta^*$,
		\begin{align*}
			V_{2b}(\theta_1 \from \theta_2) &= V_{2b}(\theta_1 \from \theta_2) - V_{2b}(\theta_1 \from \theta_1) \\
			&=\left[2V_{1b}(\theta_1 \from \theta_2)G(\theta^* - b) 
			- 2\left(\log\frac{G(\theta_1 - b)}{G(\theta^* - b)} \right)G(\theta_1 - b)\right](\theta_2 - \theta_1) \\
			&\leq \left[2 |\theta^* - \theta_1| + 2\left|\log\frac{G(\theta_1 - b)}{G(\theta^* - b)}\right|\right] |\theta_2 - \theta_1| \\
			&= \left[2|\theta^* - \theta_1| + 2|G(\theta_1 - b)\overline{G}(\theta^{**} - b)||\theta^* - \theta_1|\right]|\theta_2 - \theta_1| \\
			& \leq 4|\theta_2 - \theta_1|^2.
		\end{align*}
		Next we show that $h_b^2(\theta_1, \theta_2) \leq 4|\theta_1 - \theta_2|^2$ for any $\theta_1$, $\theta_2$ in $\R$.
		Define for any $x, y, b \in \Theta$,
		\begin{align*}
			l_b(x, y) = \overline{G}(y - b)[G(y - b) - G(y - b)^{\frac{1}{2}}G(x - b)^{\frac{1}{2}}] \\
			m_b(x, y) = G(y - b)[\overline{G}(y - b)^{\frac{1}{2}}\overline{G}(x - b)^{\frac{1}{2}} - \overline{G}(y - b)].
		\end{align*}
		By computation, 
		\begin{align*}
			\vpd{}{\theta_2}h_b^2(\theta_1, \theta_2) &= l(\theta_1, \theta_2) + m(\theta_1, \theta_2).\\
			\vpd{}{x}l_b(x, y) &= \frac{-1}{2} \left[G(x - b)G(y - b)\right]^{\frac{1}{2}} \overline{G}(y - b) \overline{G}(x - b).\\
			\vpd{}{x}m_b(x, y) &= \frac{-1}{2} \left[\overline{G}(x - b)\overline{G}(y - b)\right]^{\frac{1}{2}} G(y - b) G(x - b).
		\end{align*}
		Note that $h_b^2(\theta, \theta) = 0$, $l_b(\theta, \theta) = m_b(\theta, \theta) = 0$ for any $\theta \in \Theta$.
		By Mean Value theorem, for some $\theta^*$ between $\theta_1$ and $\theta_2$, 
		$\theta^{**}$ between $\theta_1$ and $\theta^*$,
		\begin{align*}
			h_b^2(\theta_1, \theta_2) &= \left| l_b(\theta_1, \theta^*) + m_b(\theta_1, \theta^*)\right||\theta_1 - \theta_2| \\
			&= \left| \vpd{l_b(\theta^{**}, \theta^*)}{x} + \vpd{m_b(\theta^{**}, \theta^*)}{x}\right||\theta^* - \theta_1||\theta_1 - \theta_2| \\
			& \leq |\theta_1 - \theta_2|^2.
		\end{align*}
		
		Next, we prove that there exists such $c > 0$.
		It suffices to prove that the function $(\theta_1, \theta_2, b) \mapsto |\theta_1 - \theta_2|^2 / h_b^2(\theta_1, \theta_2)$
		is bounded above by a constant.
		Denote $A = \Theta \bigcup \{b_k\}$.
		Since $\Theta$ and $\{b_k\}$ is bounded, we only have to check whether the function is bounded above 
		by some constant when $h_b^2(\theta_1, \theta_2)$ is close to $0$ (i.e. when $\theta_1$ and $\theta_2$ are close).
		Fix $\theta_1$ and $b$. By L'Hospital's theorem, 
		\begin{align*}
			\lim_{\theta_2 \to \theta_1} \frac{(\theta_2^2 - \theta_1^2)}{h_b^2(\theta_1, \theta_2)} 
			&=\lim_{\theta_2 \to \theta_1} \frac{2 \theta_2}{l_b(\theta_1, \theta_2) + m_b(\theta_1, \theta_2)} \\
			&=\lim_{\theta_2 \to \theta_1} \frac{2(\theta_2 - \theta_1)}{\left( \vpd{l_b(\theta^*, \theta_2)}{x} + \vpd{m_b(\theta^*, \theta_2)}{x}\right)(\theta_1 - \theta_2)} \\
			&= 2\left| \vpd{l_b(\theta_1, \theta_1)}{x} + \vpd{m_b(\theta_1, \theta_1)}{x}\right|^{-1}.
		\end{align*}
		Let $m_1 = \inf_{\theta, b \in A} G(\theta - b) > 0$ and $m_0 = \inf_{\theta, b \in A} \overline{G}(\theta - b) > 0$.
		Then $\left| \vpd{l_b(\theta_1, \theta_1)}{x} + \vpd{m_b(\theta_1, \theta_1)}{x}\right| \geq \frac{1}{2}m_0m_1(m_0 + m_1)$.
		Therefore,
		\[ \lim_{\theta_2 \to \theta_1} \frac{(\theta_2^2 - \theta_1^2)}{h_b^2(\theta_1, \theta_2)} < 4[m_0m_1(m_0 + m_1)]^{-1} < \infty.\]
	}
\end{proof}
\newpage

\end{document}